\documentclass[pre,twocolumn,citesort,aps,floats]{revtex4}%
\usepackage{amsfonts}
\usepackage{amsmath}
\usepackage{amssymb}
\usepackage{graphicx}%
\providecommand{\U}[1]{\protect\rule{.1in}{.1in}}
\providecommand{\U}[1]{\protect\rule{.1in}{.1in}}
\providecommand{\U}[1]{\protect\rule{.1in}{.1in}}
\providecommand{\U}[1]{\protect\rule{.1in}{.1in}}
\begin{document}
\title{Induced-charge electrophoresis near an insulating wall}
\author{Mustafa Sabri Kilic$^{1}$ and Martin Z. Bazant$^{1,2}$}
\affiliation{$^{1}$ Department of Mathematics, Massachusetts Institute of Technology,
Cambridge, MA 02139,USA}
\affiliation{$^{2}$ Physico-Chimie Th\'eorique, Gulliver-CNRS, ESPCI, 10 rue Vauquelin,
Paris 75005, France}

\begin{abstract}
Induced-charge electrophoresis (ICEP) has mostly been analyzed for asymmetric
particles in an infinite fluid, but channel walls in real systems further
break symmetry and lead to dielectrophoresis (DEP) in local field gradients.
Zhao and Bau (Langmuir, \textbf{23}, 2007, pp 4053) recently predicted that a
metal (ideally polarizable) cylinder is repelled from an insulating wall in a
DC field. We revisit this problem with an AC field and show that attraction to
the wall sets in at high frequency and leads to an equilibrium distance, where
DEP balances ICEP, although, in three dimensions, a metal sphere is repelled
from the wall at all frequencies. This conclusion, however, does not apply to
asymmetric particles. Consistent with the recent experiments of Gangwal et al.
(arXiv:0708.2417), we show that a metal/insulator Janus particle is always
attracted to the wall in an AC field. The Janus particle tends to move toward
its insulating end, perpendicular to the field, but ICEP torque rotates this
end toward the wall. Under some conditions, the theory predicts steady
translation along the wall with an equilibrium tilt angle, as seen in
experiments, although more detailed modeling of the contact region of
overlapping double layers is required.

\end{abstract}
\date{\today }
\maketitle

\centerline{ DRAFT }

\section{Introduction}

Most theoretical work on electrophoresis has focused on spherical particles
moving in an infinite fluid in response to a uniform applied electric
field~\cite{anderson1989,lyklema_book_vol2,hunter_book,russel_book}. Of
course, experiments always involve finite geometries, and in some cases walls
play a crucial role in electrophoresis. The linear electrophoretic motion of
symmetric (spherical or cylindrical) particles near insulating or dielectric
walls~\cite{morrison1970,keh1985,keh1988,keh1991,keh1996,ennis1997} and in
bounded cavities or
channels~\cite{zydney1995,lee1997,lee1998,shugai1999,chih2002,liu2004,hsu2005,davison2006,hsu2007b,hsu2007c}
has been analyzed extensively. Depending on the geometry and the double-layer
thickness, walls can either reduce or enhance the translational velocity, and
the rotational velocity can be opposite to the rolling typical of sedimention
near a wall. The classical analysis for thin double layers assumes
``force-free'' motion driven by electro-osmotic slip alone, but recent work
has shown that electrostatic forces can also be important near
walls~\cite{yariv2006,hsu2007a}. Heterogeneous particles with non-uniform
shape and/or zeta potential exhibit more complicated bulk
motion~\cite{anderson1984,fair1992,long1996,long1998}, which can also affect
boundary interactions~\cite{tang2001}, especially if the particles are
deformable, as in the case of chain-like biological
molecules~\cite{randall2005}.

In this article, we focus on the effect of nonlinear induced-charge
electro-osmotic (ICEO) flows at polarizable surfaces, which are finding many
new applications in microfluidics and
colloids~\cite{iceo2004a,iceo2004b,squires2005}. The canonical example of
quadrupolar ICEO flow around a polarizable particle, first described by
Murtsovkin~\cite{murtsovkin1996,gamayunov1986}, involves fluid drawn in along
field axis and expelled radially in the equatorial plane in an AC or DC field.
Broken symmetries in this problem can generally lead to hydrodynamic forces
and motion induced-charge electrophoresis (ICEP), as well as electrical forces
and motion by dielectrophoresis (DEP). Such phenomena have only been analyzed
for isolated asymmetric particles in an infinite
fluid~\cite{iceo2004a,yariv2005,squires2006} or in a dilute solution far from
the walls~\cite{saintillan2006,rose2007}. In contrast, experiments
demonstrating translational ICEP motion have involved strong interactions with
walls~\cite{murtsovkin1990,velev}, which remain to be explained.%

\begin{figure}
[ptb]
\begin{center}
\includegraphics[
trim=0.000000in 0.000000in -0.002333in 0.000000in,
height=1.7071in,
width=3.0407in
]%
{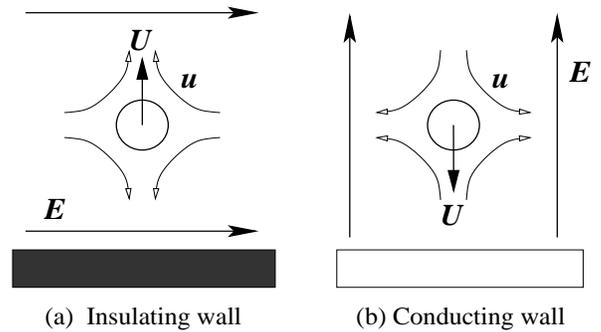}%
\caption{ Hydrodynamic forces on polarizable particles near (a) insulating and
(b) unscreened conducting walls due to ICEO flows}%
\label{fig:wall}%
\end{center}
\end{figure}

As shown in Figure~\ref{fig:wall}, it is easy to see that the quadrupolar ICEO
flow around a polarizable particle typically causes attraction to unscreened
conducting walls (perpendicular to the field) and repulsion from insulating
walls (parallel to the field). The former effect of ICEP attraction to
conducting walls has not yet been analyzed; it may play a role in colloidal
self assembly on electrodes applying AC voltages
~\cite{trau1997,yeh1997,sides2001,ristenpart2003,ristenpart2007}. This
phenomenon is mainly understood in terms of electrohydrodynamic flows (what we
would term \textquotedblleft ICEO\textquotedblright) induced on the
electrodes, not the particles (typically latex spheres), but ICEP could be
important for more polarizable particles.

The latter effect of ICEP repulsion from insulating walls has recently been
analyzed by Zhao and Bau~\cite{zhao2007} in the case of a two-dimensional
ideally polarizable cylinder in a DC field. However, this phenomenon has not
yet been confirmed experimentally. On the contrary, Gangwal et al~\cite{velev}
have recently observed that metallo-dielectric Janus particles are attracted
to a glass wall, while undergoing ICEP motion parallel to the wall and
perpendicular to an applied AC field. It is not clear that the existing theory
of ICEP can explain this surprising behavior.

The objective of this work is to analyze the motion of three-dimensional
polarizable particles near insulating walls in AC fields. As summarized in
section ~\ref{sec:model}, we employ the standard low-voltage model in the thin
double-layer approximation, following many
authors~\cite{encyclopedia_polarizable,iceo2004b,levitan2005,yariv2005,squires2006}%
, including Zhao and Bau~\cite{zhao2007}. In section \ref{sec:cyl}, we first
analyze ideally polarizable cylinders and spheres near a non-polarizable wall,
which only experience forces perpendicular to the wall. In section
\ref{sec:janus} we then study spherical metal/insulator Janus particles, which
are half ideally polarizable and half non-polarizable. Due to their broken
symmetry, the Janus particles also experience ICEP and DEP torques, which
strongly affect their dynamics near the wall.

\section{Mathematical Model}

\label{sec:model}

\subsection{Low Voltage Theory}

In this paper, we will consider either a cylindrical or a spherical particle
of radius $a$ in a semi-infinite electrolyte bounded by a plane. The distance
between the center of the particle and the plane is denoted by $h.$ In the
absence an applied electric field, we assume that the particle and the wall
surfaces are uncharged. In addition, we will assume the electrolyte has a low
Reynolds number, and impose Stokes equations. We will assume that the thin
double layer approximation holds and the bulk electrolyte remains
electroneutral, which is the case when the Debye length%
\[
\lambda_{D}=\sqrt{\frac{\varepsilon kT}{2z^{2}e^{2}c_{0}}}%
\]
is much smaller than the characteristic length scale (in our case, $a$). The
Debye length is typically ranges between $1-100nm,$ and colloidal particles
are usually in the $\mu$ range, therefore thin double layer approximation
holds for most of the time.

Then the general equations consist of the Laplace's
\[
\varepsilon\nabla^{2}\phi=0
\]
and Stokes equations%
\begin{align}
\eta\nabla^{2}\mathbf{u}  &  =\nabla p\nonumber\\
\nabla\cdot\mathbf{u}  &  =0 \label{stokes}%
\end{align}
where $\phi$ is the electrostatic potential and $\varepsilon$ the
permittivity, $\eta$ the viscosity of the electrolyte, $\mathbf{u}$ the
velocity field and $p$ the pressure. The wall boundary $z=0$ is an insulator,
satisfying%
\[
\mathbf{n}\cdot\mathbf{\nabla}\phi=0
\]
whereas the particle surface, being polarizable, acts as a capacitor in the
thin double layer limit%
\[
\frac{dq}{dt}=(-\mathbf{n)}\cdot(-\kappa\mathbf{\nabla}\phi)
\]
with $q$ being the surface charge density on the particle, $\kappa$ the
conductivity of the bulk electrolyte. Far away from the particle, an electric
field
\[
\mathbf{\nabla}\phi\sim\mathbf{E}_{\infty}=E_{\infty}\mathbf{\hat{x},}\text{
}|\mathbf{x}|\rightarrow\infty
\]
is applied. In general, the magnitude $E_{\infty}=E_{\infty}(t)$ may be time dependent.

The Stokes equations are supplied by the no-slip conditions on the wall and
the Smoluchowski's electrokinetic slip formula on the particle%
\[
\mathbf{u=u}_{slip}\mathbf{=}\frac{\varepsilon}{\eta}\zeta\mathbf{\nabla}%
_{s}\phi
\]
where $\zeta$ is the potential difference between the surface and the bulk.
Later, we will take into account the steric effects of ions and show how this
formula needs to be modified. Lastly, we assume flow vanishes at the infinities.

So far, the equations are complete except for a constitutive relation between
$\zeta$ and $q.$ The linear theory asserts that
\[
q=-\frac{\varepsilon}{\lambda_{D}}\zeta
\]
although the complete problem is still nonlinear in this case because of the
quadratic slip formula (\ref{slipn}). In this paper, we will study only the
linear theory, but the calculations can be repeated with (sometimes more
accurate) nonlinear theories like that of Poisson-Boltzmann.

\subsection{Force and Torque on the Particle}

Throughout this paper, we will assume that the particle is fixed and calculate
the forces on the particle. For the case of a moving particle, the slip
velocity needs to be modified to account for the motion of the particle surface.

The total force and torque on any volume of the fluid are conveniently given
in terms of the stress tensor, $\mathbf{\sigma,}$ by
\begin{align}
\mathbf{F}  &  =\int_{\partial\Omega}\mathbf{n\cdot\sigma}dA\label{force}\\
\mathbf{T}  &  =\int_{\partial\Omega}\mathbf{r}\times(\mathbf{n\cdot\sigma)}dA
\label{torque}%
\end{align}
The stress tensor contains from electrical and viscous stresses on the fluid,
$\mathbf{\sigma=\sigma}_{M}+\mathbf{\sigma}_{H},$ where%
\begin{align*}
\mathbf{\sigma}_{M}  &  =\varepsilon\lbrack\mathbf{EE}-\frac{1}{2}%
E^{2}\mathbf{I]}\\
\mathbf{\sigma}_{H}  &  =-p\mathbf{I}+\eta\left(  \mathbf{\nabla
u}+(\mathbf{\nabla u})^{T}\right)
\end{align*}
are the Maxwell and hydrodynamic stress tensors, respectively.

\subsection{Particle Dynamics}

In order to calculate the movement of a colloidal particle, we need to find a
translational velocity $\mathbf{U},$ and a rotational velocity $\mathbf{\Omega
}$ such that the net force on the particle is zero, when the slip velocity is
modified by taking into account the velocities $\mathbf{U}$ and
$\mathbf{\Omega}.$ In other words, we are seeking $\mathbf{U}$ and
$\mathbf{\Omega}$ such that the problem (\ref{stokes}) with boundary
condition
\[
\mathbf{u=u}_{slip}\mathbf{+U+r\times\Omega}%
\]
yields $\mathbf{F=0}$ and $\mathbf{T=0.}$

Since Stokes problem is linear, there is a linear relationship between the
translational and rotational motion of the particle and the resulting force
and torque exerted on it by the fluid. Let us denote this relationship by
\[
\left(
\begin{array}
[c]{c}%
\mathbf{F}\\
\mathbf{T}%
\end{array}
\right)  =\mathbf{M}\left(
\begin{array}
[c]{c}%
\mathbf{U}\\
\mathbf{\Omega}%
\end{array}
\right)
\]
The viscous hydrodynamic tensor $M$ comes from solving for the Stokes flow
around a particle moving with translational velocity $\mathbf{U}$ and and
rotational velocity $\mathbf{\Omega}$, assuming no slip on all particle and
wall surfaces.

If we then solve the electrokinetic problem for a fixed particle in the
applied field, we obtain the ICEO slip velocity $\mathbf{u}_{slip}$ as well as
the total (hydrodynamic $+$ electrostatic) force $\mathbf{F}_{slip}$ and
torque $\mathbf{T}_{slip}$ needed to hold the particle in place, thereby
preventing ICEP and DEP motion.

Using these calculations and invoking linearity, the condition of zero total
force and torque on the particle, $\left(
\begin{array}
[c]{c}%
\mathbf{F}\\
\mathbf{T}%
\end{array}
\right)  +\left(
\begin{array}
[c]{c}%
\mathbf{F}_{slip}\\
\mathbf{T}_{slip}%
\end{array}
\right)  =0,$ determines the motion of the particle
\begin{equation}
\left(
\begin{array}
[c]{c}%
\mathbf{U}\\
\mathbf{\Omega}%
\end{array}
\right)  =-\mathbf{M}^{-1}\left(
\begin{array}
[c]{c}%
\mathbf{F}_{slip}\\
\mathbf{T}_{slip}%
\end{array}
\right)  \label{eq:force-slip}%
\end{equation}
The particle trajectory is then described by the solution to the differential
equation%
\[
\frac{d\mathbf{x}}{dt}=\mathbf{U}%
\]
together with the equations for the particle's angular orientation.

This angular orientation does not matter for a full polarizable or insulating
particle. For the Janus particle, we will argue that only the rotations about
$x-$axis are important, thus we will focus on the dynamics of just a single
angle. In this case, the equation of motion is simply%
\[
\frac{d\theta}{dt}=\Omega_{x}%
\]

\subsection{Nondimensional Equations}

We nondimensionalize the variables by
\begin{align*}
\mathbf{x}^{\prime}  &  =\frac{\mathbf{x}}{a},\text{\quad}\phi^{\prime}%
=\frac{\phi}{E_{\infty}a},\quad\zeta^{\prime}=\frac{\zeta}{E_{\infty}a}\\
q^{\prime}  &  =\frac{\varepsilon E_{\infty}a}{\lambda_{D}}q\\
t^{\prime}  &  =\left(  \frac{\lambda_{D}a}{D}\right)  ^{-1}t,\text{
}t^{\prime\prime}=\left(  \frac{\eta}{\varepsilon E_{\infty}^{2}}\right)
^{-1}t\\
\mathbf{u}^{\prime}  &  \mathbf{=}\mathbf{u}\left(  \frac{\varepsilon
E_{\infty}^{2}a}{\eta}\right)  ^{-1},\text{\qquad}p^{\prime}=\frac
{p}{\varepsilon E_{\infty}^{2}}%
\end{align*}
Note that there are two time scales in the problem, $\tau^{\prime}=$
$\frac{\lambda_{D}a}{D},$ the charging time, and $\tau^{\prime\prime}%
=\frac{\eta}{\varepsilon E_{\infty}^{2}},$ the time scale for particle motion.

\bigskip Plugging in the equations, we obtain (after dropping the primes
except for $t$)%
\begin{align*}
\nabla^{2}\phi &  =0\\
\nabla^{2}\mathbf{u}  &  =\nabla p\\
\nabla\cdot\mathbf{u}  &  =0
\end{align*}
with the boundary conditions%
\begin{align}
\frac{dq}{dt^{\prime}}  &  =\mathbf{n}\cdot\mathbf{\nabla}\phi\nonumber\\
\mathbf{u}\text{ }  &  \mathbf{=}\text{ }\zeta\mathbf{\nabla}_{s}%
\phi\label{slipn}%
\end{align}
on the particle surface, where $\zeta=\phi_{surface}-\phi_{bulk},$ is the zeta
potential. For a polarizable particle, we have $\phi_{surface}=0$ by symmetry,
therefore we are left with $\zeta=-\phi_{bulk}=-\phi.$

In addition, we have
\[
\mathbf{\nabla}\phi\sim\mathbf{\hat{x},}\text{ }|\mathbf{x}|\rightarrow\infty
\]
just as before, now with the dimensionless variables. The remaining equations
are the no-slip boundary condition on the planar wall, and the requirement
that the flow vanishes at infinities.

The constitutive relation between $q$ and $\zeta$ takes the simple form
\[
q=-\zeta=\phi
\]
for the linear theory that we are going to analyze in this paper.

The dimensionless force and torque on the particle are given by the formulae
(\ref{force}) and (\ref{torque}), where the stress tensors are replaced by
their dimensionless counterparts
\begin{align*}
\mathbf{\sigma}_{M}  &  =\mathbf{EE}-\frac{1}{2}E^{2}\mathbf{I}\\
\mathbf{\sigma}_{H}  &  =-p\mathbf{I}+\left(  \mathbf{\nabla u}%
+(\mathbf{\nabla u})^{T}\right)
\end{align*}
The force, angular momentum and stress tensors are scaled to
\[
F_{ref}=\varepsilon E_{\infty}^{2}a,\text{ }T_{ref}=\varepsilon E_{\infty}%
^{2}a^{2},\text{ }\sigma_{ref}=\varepsilon E_{\infty}^{2}%
\]
Finally, the particle motion will be governed by
\[
\frac{d\mathbf{x}}{dt^{\prime\prime}}=\mathbf{U,}\text{ }\frac{d\theta
}{dt^{\prime\prime}}=\Omega_{x}%
\]

\subsection{Simplifications}

\subsubsection{Steady Problems}

If a DC voltage is applied, then the system reaches a steady state after a
while and the time derivatives drop out. This is the case when Neumann
boundary conditions are valid also on the cylinder or sphere. In that case
\begin{align*}
F_{E}  &  =\int_{\partial\Omega}(\mathbf{EE}-\frac{1}{2}E^{2}\mathbf{I)n}dA\\
&  =\int_{\partial\Omega}(\mathbf{EE\cdot n}-\frac{1}{2}E^{2}\mathbf{n)}%
dA=-\frac{1}{2}\int_{\partial\Omega}E^{2}\mathbf{n}dA
\end{align*}
because $\mathbf{E\cdot n=}0$ on the surface. As a consequence, the
electrostatic torque induced on the particle is zero.

\subsubsection{Symmetry}

For the full cylinder problem, the electrostatic problem has an odd symmetry
in $x-$direction, that is
\[
\phi(x,z)=-\phi(-x,z)
\]
and for the full sphere problem, it has an odd symmetry in $x-$direction and
an even symmetry in $y$-direction:%
\[
\phi(x,y,z)=\phi(x,\pm y,z)=-\phi(-x,y,z)
\]
As a result, $E^{2}$ has even symmetry in $x$ and $y.$ Therefore, in the
steady case, the electrostatic forces vanish in those directions, and there
can only be a vertical force. In general time-dependent cases, $E$ has an even
symmetry in $x$ and and odd symmetry in $y,$ and therefore the electrostatic
force may not vanish in $x$ direction.

As for the Stokes problem, a glance at the slip formula shows that the slip,
just like the potential field, has an odd symmetry in both $x$ and $y,$ and
therefore so does the flow field. Consequently, there are no hydrodynamic
forces in those directions.

Needless to say, all these symmetry arguments disappear for the Janus particles.

\subsubsection{AC fields}

\bigskip In the linear model we are considering, the time-periodic
electrostatic forcing problem can be solved by letting
\[
\phi=\operatorname{Re}(\tilde{\phi}e^{i\omega t})
\]
and solving for the complex potential $\tilde{\phi}$ using the equations%
\[
\nabla^{2}\tilde{\phi}=0
\]
with the boundary conditions%

\begin{align*}
n\cdot\nabla\tilde{\phi}  &  =i\omega\tilde{\phi}\text{ (polarizable)}\\
n\cdot\nabla\tilde{\phi}  &  =0\text{ (insulator)}\\
\tilde{\phi}  &  =-E_{\infty}x\text{ (at infinity)}%
\end{align*}

In the high frequency limit, the electrostatic problem approaches to the
solution of the dirichlet problem, that is, the first boundary condition is
replaced by
\begin{equation}
\tilde{\phi}=0\text{ (polarizable)} \label{dirichlet}%
\end{equation}
This is because $\tilde{\phi}=n\cdot\nabla\tilde{\phi}/i\omega\approx0.$
Physically, this means that the double layers do not have enough time to
charge when the forcing frequency is too high.

Note that in cases of AC fields below, it is the field amplitude $E_{0}$ that
enters into the scalings above, and the $U$ is the time-averaged velocity.

Once the electrostatic potential is calculated, the time-averaged slip
velocity can be obtained by the formula%
\begin{equation}
\mathbf{u}_{s}=\frac{1}{2}\operatorname{Re}[\tilde{\zeta}\tilde{E}_{//}^{\ast
}] \label{uscomplex}%
\end{equation}
where $\tilde{\zeta}$ is the (complex) surface zeta potential, which is equal
to $-\tilde{\phi}$ in the linear theory, and $\tilde{E}_{//}^{\ast}$ is the
complex conjugate of the tangential component of $\tilde{E}=\nabla\tilde{\phi
}$, the complex electric field.

In the DC limit as $\omega\rightarrow0,$ the imaginary parts of the solutions
go to zero, and we are left with $\mathbf{u}_{s}=\frac{1}{2}\zeta E_{//},$
which is the standard Smoluchowski's formula with a factor of $1/2$.

\subsection{Numerical Methods}

We have solved the equations using the finite element software COMSOL for
various geometries which have been approximated by the cylindrical or
spherical colloidal particle being enclosed by a finite rectangle. The
equations are first converted to their weak forms, and entered into the
general weak PDE module of COMSOL. None of the COMSOLs special modules are
used.\bigskip%
\begin{figure}
[ptb]
\begin{center}
\includegraphics[
height=2.5936in,
width=3.4411in
]%
{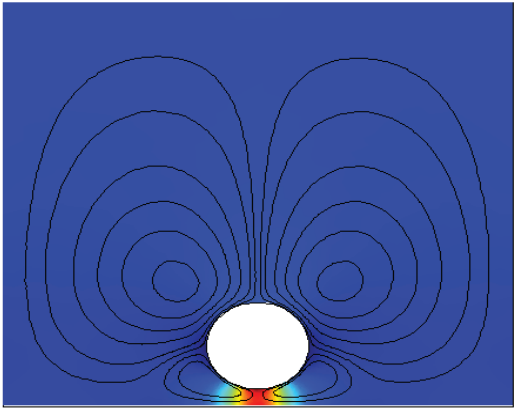}%
\caption{Streamlines to the Stokes flow problem for a cylinder near a wall.
The surface color indicates the pressure. In this case, the increased pressure
between the wall and the cylinder leads to repulsion away from the wall. }%
\end{center}
\end{figure}

For linear and nonlinear models alike, the computational efficiency is
improved by first solving the electrostatic problem, and then the hydrodynamic
problem. In time dependent cases, the fluid slip can be averaged and the
Stokes problem is solved only once using this averaged slip.

For reasons of completeness and easy reference, here we list the weak forms of
the equations solved. This system of equations are converted to weak form by
multiplying by corresponding test functions and integrating over the spatial domain.

The electrical problem turns into%
\begin{align*}
0  &  =-\int_{\Omega}\hat{\phi}\mathbf{\nabla}^{2}\phi d\mathbf{r=}%
\int_{\Omega}\mathbf{\nabla}\hat{\phi}\mathbf{\nabla}\phi d\mathbf{r+}%
\int_{\partial\Omega}\hat{\phi}(\mathbf{n}\cdot\mathbf{\nabla}\phi)ds\\
&  =\int_{\Omega}\mathbf{\nabla}\hat{\phi}\mathbf{\nabla}\phi d\mathbf{r+}%
\int_{\partial\Omega}\hat{\phi}\partial_{t}qd\mathbf{r}%
\end{align*}
which is satisfied for all test functions $\hat{\phi}.$ The boundary condition
for $\phi$ is imposed in the form
\[
0=\int\hat{q}\left(  V_{\operatorname{col}}-\phi-\zeta\right)  d\mathbf{r}%
\]
to be satisfied for all test functions $\hat{q}.$

The weak form for the stokes flow is similarly obtained as%
\begin{align*}
0  &  =-\int_{\Omega}[\mathbf{\hat{u}\cdot(\nabla\cdot\sigma)}d\mathbf{r+}%
\hat{p}\mathbf{\nabla\cdot u]}d\mathbf{r}\\
&  =-\int_{\Omega}[\mathbf{\nabla\hat{u}:\sigma-}\hat{p}\mathbf{\nabla\cdot
u]}d\mathbf{r+}\int_{\partial\Omega}\mathbf{\hat{u}\cdot}\left(
\mathbf{n\cdot\sigma}\right)  ds
\end{align*}
Since we do not have a simple expression for $\mathbf{n\cdot\sigma,}$ it is
best to introduce the new variable (Lagrangian multiplier) $\mathbf{f}%
=\mathbf{n\cdot\sigma.}$ This is also convenient for calculation of
hydrodynamic forces at the surface. Then we obtain%
\[
0=-\int_{\Omega}[\mathbf{\nabla\hat{u}:\sigma-}\hat{p}\mathbf{\nabla\cdot
u]}d\mathbf{r+}\int_{\partial\Omega}[\mathbf{\hat{u}\cdot f+\hat{f}%
\cdot(u-\mathbf{u}_{s})]}ds
\]

\section{Isotropic particles near a wall}

\label{sec:cyl}

\subsection{ Cylinder in a DC Field}

For isotropic particles near a wall, by symmetry, $\phi_{cylinder}=0,$
therefore $\zeta=-\phi$. Moreover, there is no net horizontal force exerted on
the particle, so the only forces of interest is in the vertical direction.
Another consequence of symmetry is the absence of net torque on the cylinder.

The DC cylinder problem has been solved analytically by Zhao and Bau
\cite{zhao2007} in the linear case in bipolar coordinates. The mapping between
the bipolar and the Cartesian coordinates is given by%
\[
x=\frac{c\sin\beta}{\cosh\alpha-\cos\beta},\text{ }y=\frac{c\sinh\alpha}%
{\cosh\alpha-\cos\beta}%
\]
where $\alpha_{0}<\alpha<\infty,$ and $-\pi<\beta<\pi$ defines the region
outside the cylinder. The geometric constants $\alpha_{0}$ and $c$ are defined
as
\begin{align*}
\alpha_{0}  &  =sech^{-1}(a/h)\\
c  &  =\frac{h}{\coth\alpha_{0}}%
\end{align*}
(note that there is a typo in the expression for $\alpha_{0}$ in
\cite{zhao2007}). The hydrodynamic and electrostatic forces on the cylinder
are calculated to be
\begin{align*}
F_{H}  &  =\frac{2\pi\sinh\alpha_{0}E_{\infty}^{2}c}{\left(  \alpha_{0}%
\cosh\alpha_{0}-\sinh\alpha_{0}\right)  \coth\alpha_{0}}\times\{\frac
{1}{2\sinh^{2}\alpha_{0}}+\\
&  \sum_{n=1}^{\infty}\left(  \frac{\cosh\alpha_{0}}{\sinh(n+1)\alpha_{0}%
\sinh\alpha_{0}}-\frac{1}{\sinh(n+2)\alpha_{0}\sinh n\alpha_{0}}\right)
\}\mathbf{\hat{y}}\\
F_{E}  &  =\frac{2\pi E_{\infty}^{2}h}{\coth\alpha_{0}}\sum_{n=1}^{\infty
}\left(  \frac{n^{2}}{\sinh^{2}n\alpha_{0}}-\frac{n(n+1)\cosh\alpha_{0}}{\sinh
n\alpha_{0}\sinh(n+1)\alpha_{0}}\right)  \mathbf{\hat{y}}%
\end{align*}
Because of symmetry, there is no force in the horizontal direction.

\begin{figure}
[ptb]
\begin{center}
\includegraphics[
height=3.5475in,
width=3.4402in
]%
{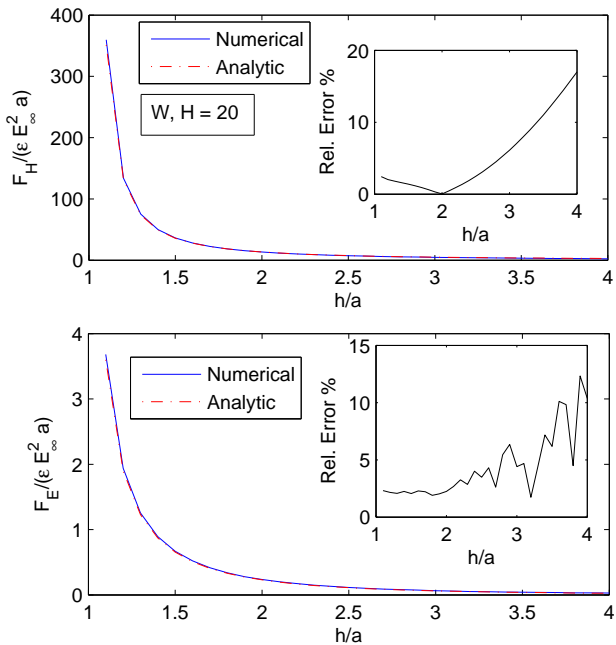}%
\caption{The Comsol numerical solution is compared to the analytical one given
by Zhao and Bau. Although the absolute errors tend to remain small, and the
curves look identical, the relative error grows fast as the particle is
located at larger distances from the planar wall. }%
\label{valid1}%
\end{center}
\end{figure}

We are going to use this solution to gain some confidence in our numerical
simulations. In Fig.\ref{valid1}, you can see the comparison of COMSOL results
with the analytical expression. The match is especially good when the particle
is close to the wall. It gets worse as this distance increases, because the
effects from the other walls also kick in. The simulation in Fig.\ref{valid1}
has run with a box of size 20x20, and maximum mesh size 1, with finer mesh on
the particle, specifically a maximum size of 0.1. Experimentation with Comsol
shows that the hydrodynamic error is sensitive to the size of the box, while
the error in the electrostatic force is more sensitive to the mesh size. For a
larger box, 40x40, and twice as finer mesh, we have obtained a similar picture
with errors cut to about one third of their values in Fig.\ref{valid1}
(results not shown).%

\subsection{ Cylinder in an AC field}

As the electric fields are screened quickly by the electrolyte, an AC field is
usually preferred. Use of an AC electric field also prevents harmful reactions
on electrodes, and enables experimentalists to go to higher applied voltage
differences. Such higher voltages may be desirable if they lead to stronger
electrokinetic effects of interest.

Far from the wall, the ICEO slip velocity around an ideally polarizable
cylinder in an AC field was derived by Squires and Bazant~\cite{iceo2004b},
which takes the dimensionless form
\begin{equation}
\langle u_{\theta}\rangle= \frac{\sin2\theta}{1 + \omega^{2}}.
\end{equation}
We use this expression to calibrate our numerical code, and find excellent
agreement far from the wall. This result shows that ICEO flow decays
algebraically as $\omega^{-2}$ above the RC charging frequency. Since
electrostatic forces do not decay in this limit, we may expect a change in
behavior near the wall. At high frequency, there is not enough time for
double-layer relaxation, so the electric field ressembles that of a conductor
in a uniform dielectric medium.%

\begin{figure}
[ptb]
\begin{center}
\includegraphics[
height=2.7657in,
width=3.4411in
]%
{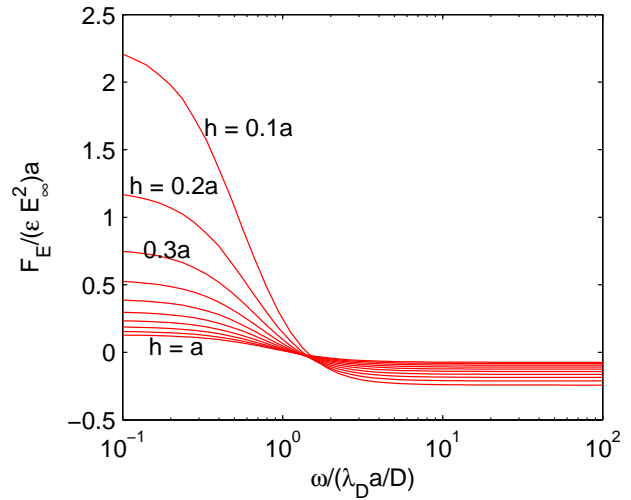}%
\caption{The total electrostatic force on the cylinder changes sign as the
frequency is increased. As the frequency approaches infinity, this force has a
finite nonzero limit. }%
\end{center}
\end{figure}
%

\begin{figure}
[ptb]
\begin{center}
\includegraphics[
height=2.4483in,
width=3.4411in
]%
{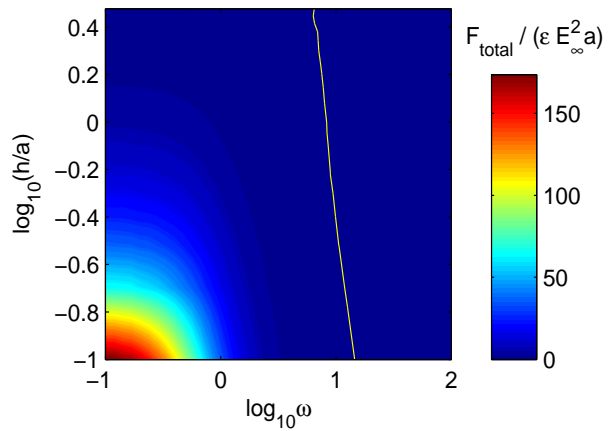}%
\caption{Contour plot of total force on the ideally polarizable cylinder.
There is an equilibrium distance between the cylinder and the wall at high
enough frequencies, indicated by the yellow contour line. As the frequency is
increased, this distance decreases.}%
\end{center}
\end{figure}

An important observation is that the total hydrodynamic forces vanishes at
higher frequencies whereas the total electrostatic force changes sign. As a
result, if the frequency is high enough, there is an equilibrium distance from
the wall. This distance decreases as the frequency is increased.

In the high frequency limit, the electrostatic problem approaches to the
solution of the dirichlet problem, that is, the laplace equation%
\[
\nabla^{2}\tilde{\phi}=0
\]
combined with the boundary condition%
\begin{align*}
\tilde{\phi}  &  =0\text{ (on the cylinder)}\\
n\cdot\nabla\tilde{\phi}  &  =0\text{ (on the wall)}\\
\tilde{\phi}  &  =-E_{\infty}x\text{ (at infinity)}%
\end{align*}
In this case, the solution is given by
\[
\phi=\operatorname{Re}(\hat{\phi}e^{i\omega t})=\hat{\phi}\cos\omega t
\]
This problem can be solved analytically, and the solution is given by
\begin{align*}
\tilde{\phi}  &  =2cE_{\infty}\sum_{n=1}^{\infty}\frac{e^{-n\alpha_{0}}}{\cosh
n\alpha_{0}}\cosh n\alpha\sin n\beta-\frac{c\sin\beta}{\cosh\alpha-\cos\beta
}\\
&  =2c\sum_{n=1}^{\infty}\left[  \frac{e^{-n\alpha_{0}}}{\cosh n\alpha_{0}%
}\cosh n\alpha-e^{-n\alpha}\right]  \sin n\beta
\end{align*}
Plugging this into the electrostatic force leads to the formula%
\begin{align*}
F_{E,\omega\rightarrow\infty}  &  =-2\pi cE_{\infty}\sum_{n=1}^{\infty}%
(\frac{n^{2}}{\cosh^{2}n\alpha_{0}}\\
&  +\frac{n(n+1)\cosh\alpha_{0}}{\sinh n\alpha_{0}\sinh\left(  n+1\right)
\alpha_{0}})
\end{align*}
with the same notation as in Zhao and Bau.

\subsection{ Sphere in an AC field}

\label{sec:sphere}

ICEO flow around a sphere was first considered by Gamayunov et
al.~\cite{gamayunov1986}. Following the cylinder analysis of Squires and
Bazant~\cite{iceo2004b}, it is straightforward to derive the (dimensionless)
ICEO slip velocity around an ideally polarizable sphere in an AC field, far
from the wall,
\begin{equation}
\langle u_{\theta}\rangle= \frac{9}{16} \frac{\sin2\theta} {1 + (\omega
/2)^{2}} \label{eq:ACsphere}%
\end{equation}
Note that since $\langle\cos^{2} \omega t \rangle= 1/2$ the ICEO flow in a
constant DC field $E_{0}$ is twice as large as the time-averaged flow in an AC
field $E_{0} \cos\omega t$ of the same amplitude: $u_{\theta}^{DC} = 2 \langle
u_{\theta}^{AC} \rangle$.

For a sphere near a wall, the results are qualitatively the same as for a
cylinder near a wall. That is, in the DC case, both the hydrodynamic and the
electrostatic forces are repulsive. Moreover, the magnitude of hydrodynamic
forces are about 2 orders of magnitude larger than the electrical forces.%

\begin{figure}
[ptb]
\begin{center}
\includegraphics[
height=1.7227in,
width=3.4411in
]%
{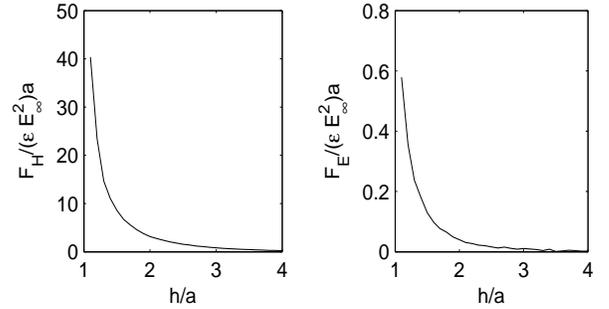}%
\caption{The hydrodynamic and electrostatic forces on a full metal sphere. }%
\end{center}
\end{figure}

Note that the steady and time-periodic plots are consistent: At $h/a=1.5,$ the
steady plots show forces about $F_{H}\approx8$ and $F_{E}\approx0.12.$ The
time-periodic plots, at the low frequency limit, start off at values
$F_{H}\approx4$ and $F_{E}\approx0.06,$ which are half of their steady counterparts.%

\begin{figure}
[ptb]
\begin{center}
\includegraphics[
height=2.0124in,
width=3.4411in
]%
{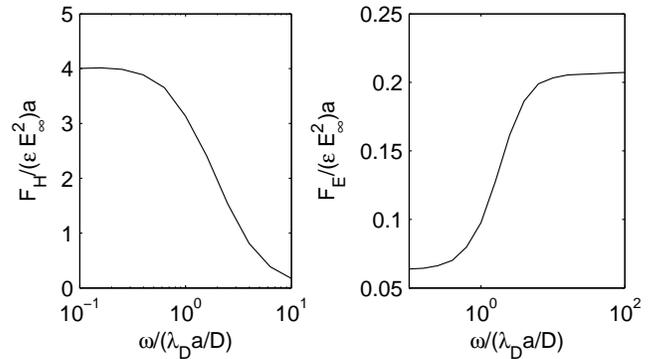}%
\caption{Both the hydrodynamic and the electrostatic forces on an
isotropic sphere are repulsive (away from the wall). While
hydrodynamic forces on the sphere decline as a function of forcing
frequency, electrostatic forces get
stronger. }%
\end{center}
\end{figure}

Unlike the cylinder problem, however, the electrostatic forces always remain
repulsive, and therefore there is no equilibrium plane for the spherical
particle, Instead, it is repelled to infinity by the wall regardless of the
forcing frequency.

\section{ Janus sphere near a wall}

\label{sec:janus}

\subsection{ Broken symmetries}

Without a nearby wall, a Janus sphere would align itself perpendicular to the
electric field, that is, some of the electric field lines would be included in
the plane dividing the Janus particles metal and insulating sides. This effect
has been demonstrated in \cite{squires2006}, and is presumably stronger than
the wall effects, at least when the particle is sufficiently away from the
wall. That being said, we will assume that the particle always stays in the
described configuration, that is, its dividing plane aligned with the electric
field. This is not to say that the particle has no room for different
rotational configurations, it can still rotate around $x$ and $y$ axis.
Rotations about the $y-$ axis (if existed) leave the particle unchanged, so we
are left with rotations only around the $x-$ axis. This is much easier to deal
with than the original problem though, as only one angle is enough the
describe the particles orientation.

Far from the wall, the bulk velocity perpendicular to a DC field in the stable
orientation is given by the formula of Squires and Bazant~\cite{squires2006}
(Eq. 3.16), which takes the dimensionless form,
\begin{equation}
U_{DC}=\frac{9}{64}=2\langle U_{AC}(\omega\rightarrow0)\rangle
\label{eq:bulkvel}%
\end{equation}
neglecting compact-layer surface capacitance. As noted above, the
time-averaged velocity in a sinusoidal AC field is smaller by a factor of two
in the limit of zero frequency. Even in the bulk, without a wall, it is
difficult to solve analytically for the ICEO flow at finite AC frequency
around a Janus particle, since the electrical response is not simply an
induced dipole, due to the broken symmetry. Nevertheless, we will see that the
frequency dependence of the flow is similar to that around a sphere
(\ref{eq:ACsphere}), constant below the RC charging time and decaying above it.

For a Janus sphere aligned perpendicular to the electric field near a wall, a
crucial observation is that the $y-$symmetry breaks down. As a result, there
is a net force in the $y-$direction, as well as a net torque in $x-$direction.
The former leads to translation parallel to the wall, while the latter causes
rotation of the dielectric face toward the wall. We shall see that these
effects of broken symmetry completely change the behavior near wall in an AC
or DC field: Although a polarizable sphere is always repelled to infinity by
an insulating wall, a Janus particle is always (eventually) attracted to it.

\subsection{ Basic mechanism for wall attraction}%

\begin{figure}
[ptb]
\begin{center}
\includegraphics[
height=1.7158in,
width=3.039in
]%
{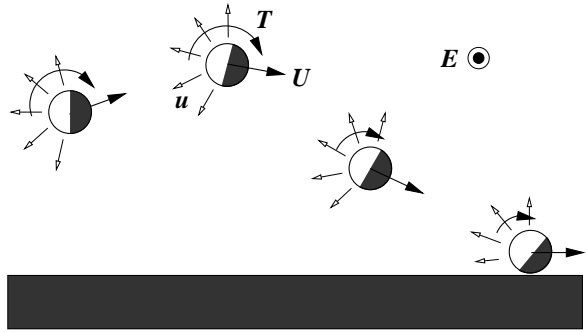}%
\caption{Sketch of ICEO flows $u$ and resulting ICEP torques $T$ which cause a
Janus particle to tilt its less polarizable end toward a wall, while
translating toward the wall (until stopped by double-layer overlap) and
perpendicular to the applied AC field $E$ (directed into the page and parallel
to the wall). This physical mechanism may explain why the transverse ICEP
motion of Janus particles was observable over the surface of a glass wall in
the experiments of Gangwal et al. ~\cite{velev}. }%
\label{fig:tilt}%
\end{center}
\end{figure}

The key new effect is rotation due to hydrodynamic torque caused by asymmetric
ICEO flow near the wall. This generally causes the Janus particle to be
attracted to the wall, as shown in figure ~\ref{fig:tilt}. The physical
mechanism can be understood as follows. When the field is first turned on, the
Janus particle quickly rotates, by ICEP and DEP, to align its metal/insulator
interface with the field axis, but with an arbitrary azimuthal angle, mainly
set by the initial condition. As described by Squires and
Bazant~\cite{squires2006}, the ICEO flow around the particle draws in fluid
along the field axis and ejects it radially at the equator -- but only on the
polarizable hemisphere, which acts like a \textquotedblleft jet
engine\textquotedblright\ drives ICEP motion in the direction of the
non-polarizable hemisphere, which leads the way like a \textquotedblleft
nose\textquotedblright.

Near a wall, as shown in the figure, the outward ICEO flow pushes down on the
wall harder on the side of the polarizable ``engine'' than on that of the
non-polarizable ``nose'', which produces a hydrodynamic torque tilting the
nose toward the wall. A second cause of this rotation is the hydrodynamic
coupling between ICEP translation parallel to the wall and rotation by shear
stresses to cause rolling past the wall. Regardless of the initial position,
these two sources of ICEP rotation cause the nose to eventually face the wall,
so that the translational engine drives it toward the wall. This is likely the
origin of the counter-intuitive attraction of Janus particles to a glass wall
in the experiments of Gangwal et al~\cite{velev}.

What happens next depends on the details of the particle-wall interaction at
very close distances. We will see that the bulk model with thin double layers
must eventually break down, since the particle eventually collides with the
wall, leading to overlapping particle and wall double layers. It is beyond the
scope of this work to accurately treat the nonlinear and time-dependent
behavior of these overlapping double layers, so we will explore two models:
(i) infinitely thin double layers, i.e. using the bulk model to arbitrarily
small heights, and (ii) a cutoff \textquotedblleft collision\textquotedblright%
\ height, where overlapping double layers stop any further motion toward the
wall, while still allowing transverse motion. The latter case assumes, as in
the experiments~\cite{velev}, that the particles and walls have
\textit{\ equilibrium} surface charge of the same sign. For concreteness, we
will simulate Model (ii) with a cutoff height $h=\lambda=0.05a$, e.g.
corresponding to a double-layer thickness (screening length) of $\lambda=50$nm
with particles of size $a=1\mu$m.

Based on the simple examples above, we expect a subtle dependence on the AC
frequency. Electrostatic DEP motion will always begin to dominate the
hydrodynamic ICEP motion at high frequency. Therefore, we now consider the low
and high frequency cases separately.
\begin{figure}
[ptb]
\begin{center}
\includegraphics[
height=3.3347in,
width=2.8971in
]%
{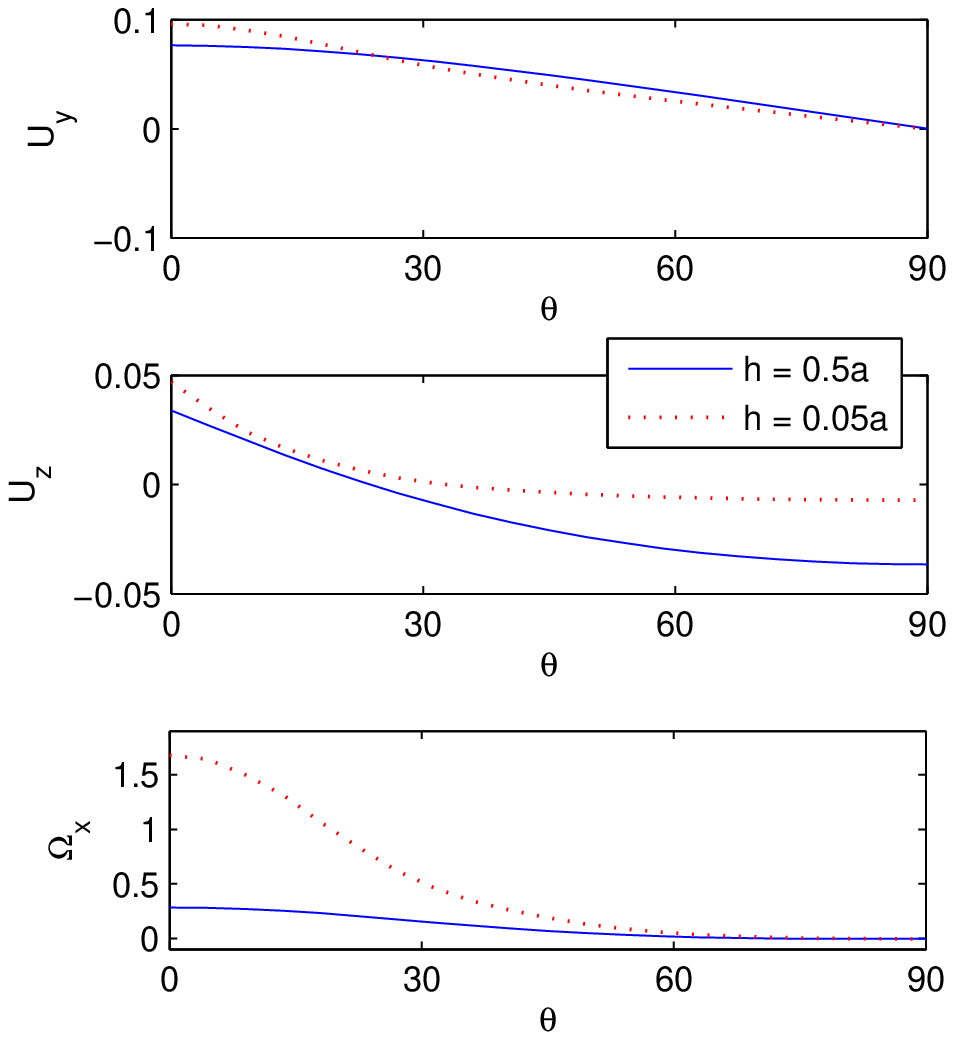}%
\caption{In the DC limit ($\omega\rightarrow0),$ we plot (a) horizontal
velocity (b) vertical velocity and (c) tilting speed (degrees/charging time)
as a function of the tilt angle $\theta$ for the janus particle at distances
$h=0.5a$ and $h=0.05a$ from the wall. }%
\label{fig:janus_DC}%
\end{center}
\end{figure}

\subsection{ Dynamics as a function of AC frequency}

As shown in Fig.~\ref{fig:janus_DC}, in the low frequency limit, the Janus
particle experiences a rotational velocity turning its non-polarizable side
toward the wall, as explained above. The hydrodynamic ICEP torque is orders of
magnitude larger than the electrostatic DEP torque, until the particle gets
quite close to the wall. The magnitude of the horizontal ICEP velocity $U_{y}$
parallel to the surface and perpendicular to the field is close to its bulk
value $U_{y}=9/128\approx0.07$ even fairly close to the wall at a height
$h=0.5a$ at zero tilt, but reduces with the tilt angle. For small tilt angles
and close to the wall at $h=0.05a$, the horizontal velocity increases to
$U_{y}\approx0.10$, but it drops below the bulk value at larger tilt angles,
e.g. to $U_{y} \approx0.05$ at $\theta=45$ degrees. Below we will see that
this velocity is further reduced at higher forcing frequencies, due to the
reduction of ICEO flow (since DEP cannot contribute to motion perpendicular to
a uniform field).%
\begin{figure}
[ptb]
\begin{center}
\includegraphics[
height=3.1687in,
width=2.6922in
]%
{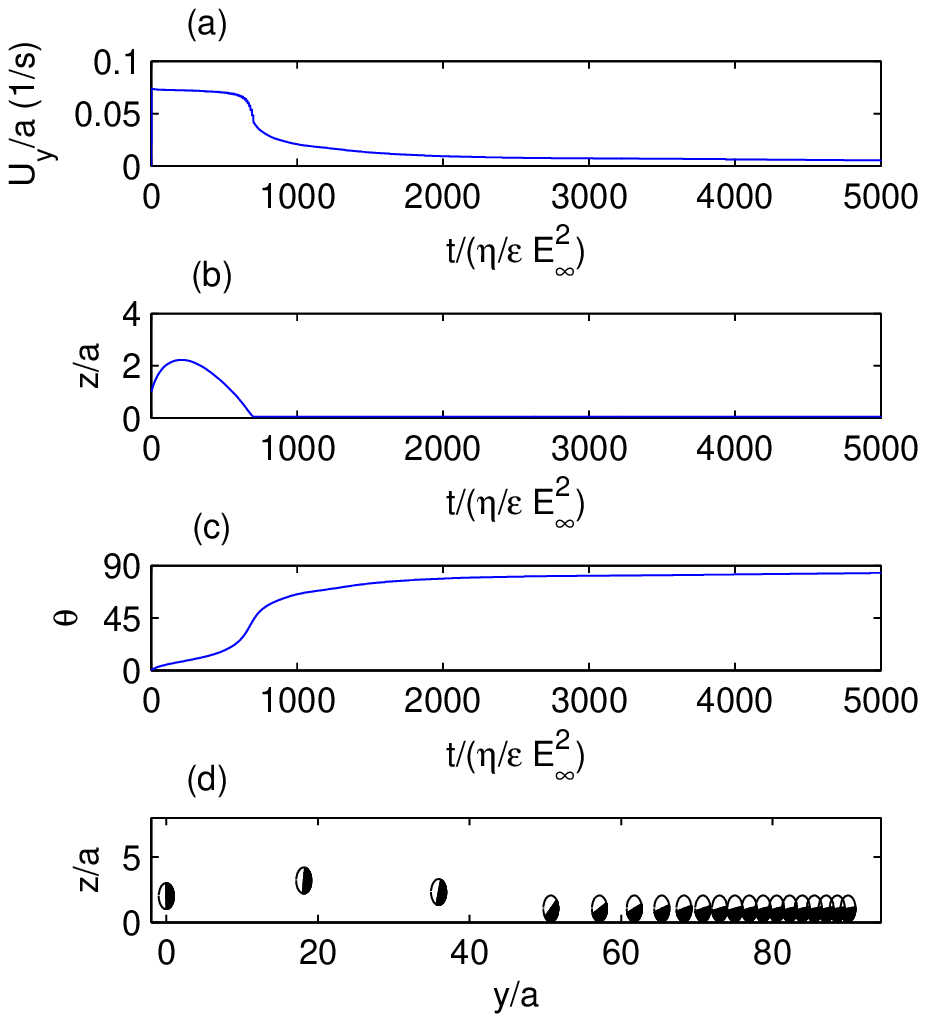}%
\caption{Typical trajectory of a janus particle under the DC limit
$\omega\rightarrow0$ interacting with the wall: As a function of time, plotted
are (a) The horizontal speed (b) Distance from the wall (c) Tilt angle. Also,
we plot the distance from the wall as a function of horizontal position in
(d).}%
\label{fig:janus_motion_DC}%
\end{center}
\end{figure}

Regardless of the orientation, in the DC limit the particle moves ever closer
to the wall in Model (i) since $U_{z}<0$ for any tilting of the nose toward
the wall. Even if the the vertical motion is stopped at a critical height in
Model (ii), the rotation continues in the DC limit until the particle points
its non-polarizable nose directly at the wall ($\theta=90$) and the motion
stops, although this can take a long time, since the rotation slows down
substantially for tilt angles larger than 45 degrees. As discussed below, a
number of effects might lead to such a stabilization of the tilt angle, thus
allowing steady translation along the wall.

As shown in Fig.~\ref{fig:janus_motion_DC}, a typical simulated trajectory of
the Janus particle shows it translating perpendicular to the field while
rotating and attracting to the wall, until eventually coming to rest facing
the wall. Even when the particle's motion stops, however, its polarizable
hemisphere (\textquotedblleft engine\textquotedblright) continues driving a
steady ICEO flow, which can lead to long-range hydrodynamic interactions with
other particles. This is an interesting theoretical prediction which should be
checked in experiments. Such immobilized Janus particles may have interesting
applications in microfluidics.%

\begin{figure}
[ptb]
\begin{center}
\includegraphics[
height=3.333in,
width=2.8115in
]%
{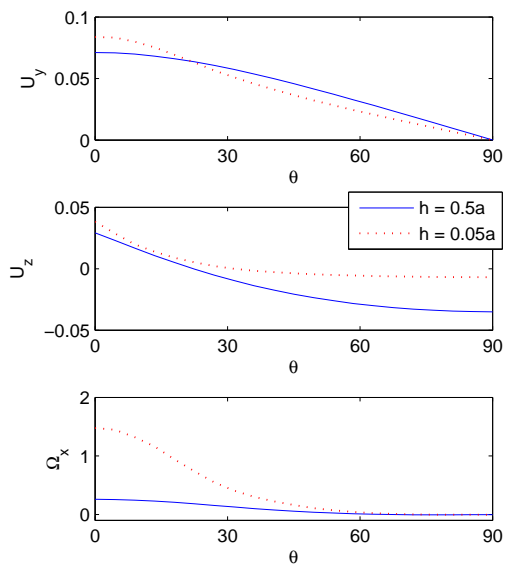}%
\caption{For AC frequency $\omega\tau_{c}=1,$ we plot (a) horizontal velocity
(b) vertical velocity and (c) tilting speed (degrees/charging time) as a
function of the tilt angle $\theta$ for the janus particle at distances
$h=0.5a$ and $h=0.05a$ from the wall. }%
\end{center}
\end{figure}

Similar behavior is predicted for finite AC frequencies in many cases. In
particular, if a particle is initially mostly facing its non-polarizable
hemisphere toward the wall ($\theta$ near 90$^{\circ}$), it will swim toward
the wall and come to rest, as in the DC limit of
Figure~\ref{fig:janus_motion_DC}.%

\begin{figure}
[ptb]
\begin{center}
\includegraphics[
height=3.1687in,
width=2.6083in
]%
{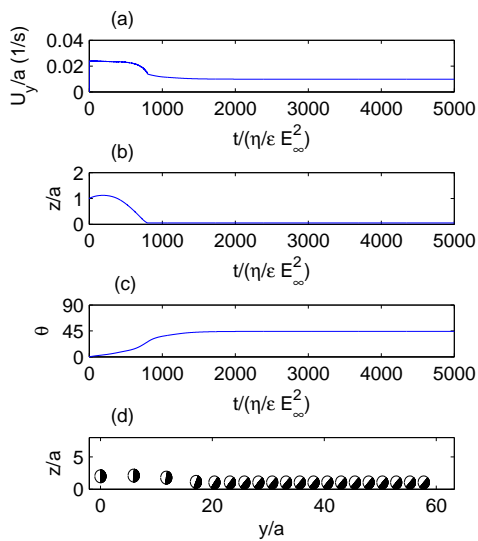}%
\caption{Typical trajectory of a janus particle under AC frequency $\omega
\tau_{c}=10$ interacting with the wall: As a function of time, plotted are (a)
The horizontal speed (b) Distance from the wall (c) Tilt angle. Also, we plot
the distance from the wall as a function of horizontal position in (d).}%
\label{fig:janus_motion_AC}%
\end{center}
\end{figure}

There are some new effects in AC fields, however, since ICEO flows are
suppressed with increasing frequency. The competing effect of DEP can prevent
the Janus particle from fully rotating and coming to rest on the surface, at
least in Model (ii) where the collision is prevented artificially, as shown in
Figure \ref{fig:janus_motion_AC}. At $\omega=1$ (the characteristic RC
frequency of the particle), the rotation slows down substantially beyond
45$^{\circ}$ but does not appear to stop. In this regime the horizontal
velocity decays to $U_{y}\approx0.015$. For $\omega=10$ the particle appears
to settles down to an equilibrium tilt angle around 45$^{\circ}$, while
steadily translating over the wall. The limiting horizontal velocity is
roughly $U_{y}\approx0.009.$%

\begin{figure}
[ptb]
\begin{center}
\includegraphics[
height=3.3347in,
width=2.8686in
]%
{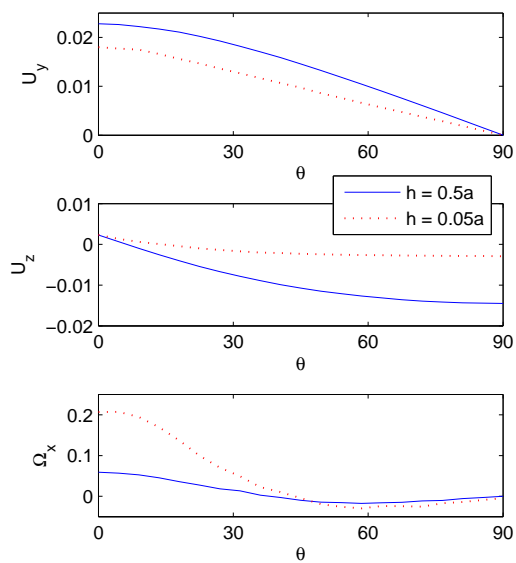}%
\caption{For AC frequency ($\omega\tau_{c}=10),$ we plot (a) horizontal
velocity (b) vertical velocity and (c) tilting speed (degrees/charging time)
as a function of the tilt angle $\theta$ for the janus particle at distances
$h=0.5a$ and $h=0.05a$ from the wall. }%
\end{center}
\end{figure}

\subsection{ Comparison to experiment}

The simulations with Model (ii) are in reasonable agreement with the
experimental observations of Gangwal et al~\cite{velev} for metallo-dielectric
Janus particles in dilute NaCl solutions in the low-frequency regime $\omega<
1$. The bulk theory of Squires and Bazant (\ref{eq:bulkvel}) accurately fits
the experimental velocity as a function of the field strength (Fig. 3 of
Ref.~\cite{velev}) and the particle size (Fig. 4), if a ratio $\delta=10$ for
the compact-layer to diffuse-layer capacitance is used to obtain the
dimensionless formula, $U_{expt} = (9/128)/(1+10) = 0.006$. However, the ICEP
motion is observed only very close to the walls.

Our simulations predict that the particles are quickly attracted to the walls
over a time of order the channel width (60 $\mu$m) divided by the typical ICEP
velocity (10 $\mu$m/s), which is roughly one minute, consistent with
experimental observations. The particles are also predicted to tilt, and
moderate tilt angles can also be inferred from experimental images, although
more accurate measurements are needed. If the tilt angle stabilizes around
45$^{\circ}$ (see below), then the simulations (Fig.~\ref{fig:janus_DC})
predict that the ICEP translational velocity should be only $0.05/0.07 = 70\%$
of the bulk value close to the wall, which would imply $\delta=7$. This value
is somewhat larger than that inferred from prior experiments on ICEO flow in
dilute KCl around a larger (100$\mu$m radius) platinum
cylinder~\cite{levitan2005}, but it is also observed that the ICEP velocity is
slower than predicted at larger sizes (Fig. 4 of Ref.~\cite{velev}). Apart
from the rotational dynamics, therefore, the theory is able to predict the
ICEP velocity fairly well.

Without stopping the rotation artificially, we are able to predict the
experimentally observed steady motion along the wall only at moderate to large
$\omega$. The reduction of ICEO flow in this regime reduces hydrodynamic
torque (see below) and also enhances the effect of stabilizing electrostatic
forces. Although $U_{expt} = 0.006$ is measured in the low-frequency plateau
$\omega< 1$, this behavior otherwise seems quite consistent, since the slower
ICEP velocity can also fit the experimental data using smaller (and perhaps
more reasonable) values of $\delta$. For example, the predicted velocity of
$U=0.015$ at $\omega=1$ implies $\delta=1.5$, while the velocity $U=0.009$ at
$\omega=10$ implies $\delta=0.5$.

The difficulty in predicting the stable tilt angle at low frequency may be due
to our use of the low-voltage, dilute-solution theory, which generally
overpredicts the magnitude of ICEO flows, especially with increasing salt
concentration. For example, the electrophoretic mobility can saturate at large
induced voltages, and the charging dynamics can also be altered significantly
when crowding effects are taken into account~\cite{large}. As a result, our
simulation results at moderate frequencies $\omega=1-10$, which exhibit
reduced ICEO flow due to incomplete double-layer charging, may ressemble the
predictions of more accurate large-voltage, concentrated-solution theories at
low frequency $\omega<1$, where flow is reduced instead by ion crowding in the
double layer. This will be the subject of future work.

\subsection{ Contact mechanics}

Another source of error in the model is our inaccurate treatment of the
contact region, where double-layers overlap. We have simply imposed a small
cutoff height in Model (ii) to prevent the wall collision, but there may be
more complicated mechanical effects of the contact region. In particular,
there may be enhanced hydrodynamic slip, due to the repulsion of overlapping
(equilibrium) double layers of the same sign, as in the experiments.%

\begin{figure}
[ptb]
\begin{center}
\includegraphics[
height=2.0799in,
width=3.4402in
]%
{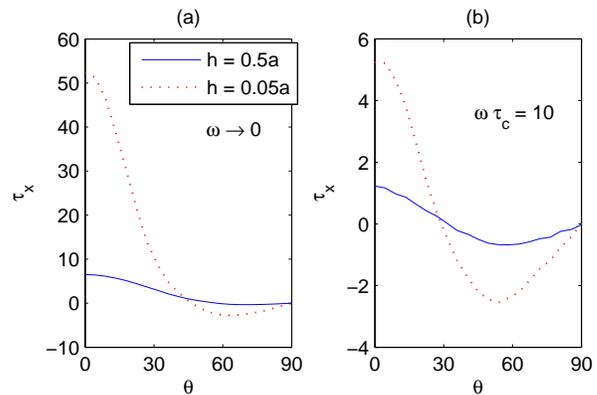}%
\caption{Torque on a fixed Janus sphere versus tilt angle at heights $h=0.5a$
and $0.05a$ when (a) $\omega\rightarrow0$ (b) $\omega\tau_{c}=10.$}%
\label{fig:janus_torque}%
\end{center}
\end{figure}
%

\begin{figure}
[ptb]
\begin{center}
\includegraphics[
height=2.0799in,
width=3.4402in
]%
{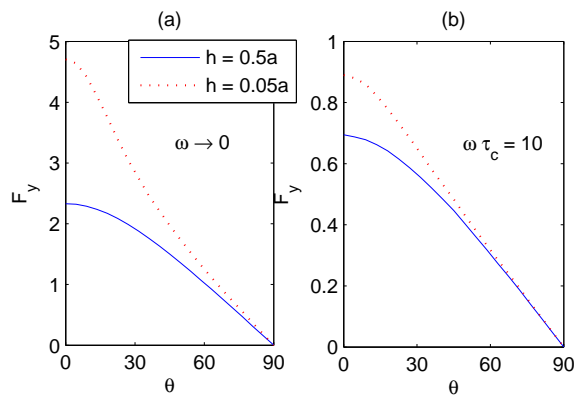}%
\caption{Horizontal force on a fixed Janus sphere versus tilt angle at heights
$h=0.5a$ and $0.05a$ when (a) $\omega\rightarrow0$ (b) $\omega\tau_{c}=10.$}%
\label{fig:janus_force}%
\end{center}
\end{figure}

By examining the forces and torques close to the wall, we can infer to some
degree what mechanical properties of the contact region might lead to the
observed ICEP sliding along the wall and smaller tilt angles at lower
frequencies (and thus also somewhat larger velocities). As shown in
Fig.~\ref{fig:janus_torque}, before the particle gets very close to the wall,
the (mostly hydrodynamic) torque acts to completely tilt the non-polarizable
face toward the wall leading to collision. As noted above in
Fig.~\ref{fig:tilt}, this can be understood as a result of the downward
component of ICEO flow on the polarizable hemisphere raising the pressure by
pushing on the wall on that side.

The situation changes when the particle gets very close to the wall. As shown
in Fig.~\ref{fig:janus_torque}, the torque changes sign at a tilt angle which
is roughly $45^{\circ}$. This again can be understood from Fig.~\ref{fig:tilt}%
, since the ICEO flow between the particle and the wall on the polarizable
side, which drives the torque, is mostly absent. It would thus seem that even
in a DC field, the particle would not rotate any farther, but this thinking
neglects the hydrodynamic coupling between translational force and rotational
velocity near the wall, Eq. (\ref{eq:force-slip}). In
Fig.~\ref{fig:janus_force}, we see that the force on the particle parallel to
the wall $F_{y}$ remains strong, and this leads to a rolling effect over the
wall due to shear stresses. For this reason, the rotational velocity persists
in Fig.~\ref{fig:janus_DC} even when the torque goes to zero in
Fig.~\ref{fig:janus_torque}.

The model assumes no slip on all non-polarizable surfaces, but this may not be
a good approximation near the contact point when double layers overlap. If the
equilibrium surface charges (or zeta potentials) on the non-polarizable
hemisphere and the wall have opposite signs, then the overlapping double
layers lead to a strong attraction, which would only stiffen the effective
contact with the surface, and thus only increase the viscous rolling effect
during motion along the surface. If the equilibrium surface charges (or zeta
potentials) have the same sign, however, as in the experiments on gold-coated
latex Janus particles near glass walls~\cite{velev}, then there is a strong
repulsion at the contact point. This repulsion stops the collision with the
wall in Model (ii), but it may also \textquotedblleft
lubricate\textquotedblright\ the contact and allow for some sliding. This
effective slip over the wall near the contact point could reduce the viscous
rolling, and, in the absence of torque, cause the rotation to stop, or at
least be reduced for tile angles above $45^{\circ}$. In that case, we might
expect a more accurate model of the contact region to predict to the
experimentally observed motion, sliding over the surface by ICEP with a small
tilt angle ($\theta<45^{\circ}$), for a wider range of conditions, including
lower AC frequency, perhaps even in the DC limit.

\section{Conclusion\label{sec:concl}}

We have use the existing low-voltage theory of ICEP to predict the motion of
polarizable particles near an insulating wall. Our results for symmetric
spheres and cylinders confirm the expected repulsion from the wall due to ICEO
flow, sketched in Figure 1(a). In the case of the cylinder we show that
attraction is also possible at high frequency, where DEP from electrostatic
forces dominates slip-driven ICEP motion.

Our results for asymmetric Janus particles reveal an unexpected attraction to
the wall by a novel mechanism illustrated in Figure ~\ref{fig:tilt}, which
involves tilting of the less polarizable face toward the wall. Once it reaches
the wall, if double-layer repulsion prevents further collision, the particle
either rotates completely and ceases to move, while driving steady ICEO flow,
or reaches an equilibrium tilt angle around 45$^{\circ}$ while steadily
translating along the surface, perpendicular to the elecric field. The latter
motion only arises at moderate frequencies in our model, above the
characteristic charging frequency for the double layers, while in experiments
it is also observed at low frequencies. More accurate models taking into
account reduced ICEO flow at large voltage in non-dilute solutions and more
accurate models of the contact region may improve the agreement with experiments.

In any case, we have shown that polarizable particles can display complex
interactions with walls due to broken symmetries in ICEO flows. Attractive and
repulsive interactions can be tuned by varying the geometry of the particles
(and the walls), as well as the AC frequency and voltage. These phenomena may
find applications in separations and self-assembly of colloids or in local
flow generation in microfluidic devices.

\end{document}